\documentstyle[11pt,moriond+,epsfig]{article}
\def\be{\begin{equation}}
\def\ee{\end{equation}}
\def\bc{\begin{center}}
\def\ec{\end{center}}
\def\bea{\begin{eqnarray}}
\def\eea{\end{eqnarray}}

\def\nn{\nonumber}
\def\lp{\lambda^{\prime\prime}}
\def\rp{$R_p$} 
\def\st1{$\tilde{t}_1$}
\def\kkb{$K^0$--$\bar{K}^0$}

\begin{document}
%
%
\vspace*{1cm}
\title{INDIRECT CONSTRAINTS ON R-PARITY VIOLATING STOP COUPLINGS}
\author{P. SLAVICH}
\address{Universit\`a di Padova and INFN, Sezione di Padova, Italy}
\maketitle\abstracts{It was recently claimed that single stop production
at the Tevatron, occurring via R-parity (and baryon number) violating 
couplings, could lead to observable signals. In this talk I present some 
results of a work in progress \cite{slavich}, showing that rare $B^+$ 
decays and \kkb\ mixing strongly constrain such a possibility.}

\vspace*{1cm}
\begin{center}
{\em Talk given at the XXXV Rencontres de Moriond,\\ 
Electroweak Interactions and Unified Theories, \\
Les Arcs, March 11-18 2000.}
\end{center}
\vspace*{1.5cm}


In the Minimal Supersymmetric Standard Model (MSSM) it is assumed that
R-parity (\rp) is conserved \cite{fayet}. While the particles of the 
Standard Model are 
even under \rp, their supersymmetric partners are odd, thus the latter can 
only be produced in pairs and they always decay into final states involving 
an odd number of supersymmetric particles. 
Although it considerably simplifies the structure of the MSSM, the 
conservation of R-parity has no firm theoretical justification (for 
a review and references on versions of the MSSM with broken R-parity,
see e.g. ref. \cite{rprev}). 
The most general \rp-violating superpotential that can be written with 
the MSSM superfields contains both lepton number and baryon number violating 
terms. Their simultaneous presence is in fact strongly limited, since 
it would induce fast proton decay. However, it is  possible to allow for
the presence of baryon number violating couplings only \cite{zwirner}, 
coming from the superpotential
\be
\label{superp}
{\cal W}_{\not{B}} = \lp_{ijk}\, \epsilon^{\alpha\beta\gamma} \,\,
U^c_{i\alpha} \,D^c_{j\beta} \,D^c_{k\gamma}\; ,
\ee
where $U^c$ and $D^c$ ($c$ denotes charge conjugation) are the chiral 
superfields associated with SU(2)-singlet antiquarks, 
in a basis where the quark masses are diagonal,
$i,j,k = 1,2,3$ are flavor indices and $\alpha,\beta,\gamma = 1,2,3$ 
are color indices. The color structure of ${\cal W}_{\not{B}}$
implies that the $\lp_{ijk}$ are antisymmetric in the last two indices, 
limiting the number of independent couplings to 9.
 
In ref. \cite{berger} the production of {\em single} top 
squarks via \rp\ violation in $p \bar{p}$ collisions at the Fermilab Tevatron 
was studied. The work was motivated by the consideration that, while the 
\rp\ and baryon number violating couplings $\lp_{1jk}$ which involve an 
up (s)quark are severely constrained by the absence of neutron-antineutron 
oscillations \cite{zwirner,goity} and double nucleon decay into kaons 
\cite{masiero}, the limits on 
the couplings  $\lp_{3jk}$ which involve a top (s)quark  are much weaker:
the authors of ref. \cite{berger} quote a 95\% confidence level upper bound
$\lp_{3jk} < 1$. If the couplings are within two orders of magnitude
from this upper limit, it turns out that the rate for the production
of a single light stop \st1\ in $p \bar{p}$ collisions at $\sqrt{S} = 2$
TeV exceeds the rate for 
stop-antistop pair production, due to the greater phase space available. 
Thus, \rp\ violation could be the favorite scenario for the observation
of supersymmetric particles at the Tevatron.

The authors of ref. \cite{berger} considered a minimal supergravity-inspired 
scenario,
where at the Grand Unified Theory (GUT) scale the common gaugino mass is 
$m_{1/2} = 150$ GeV, the scalar trilinear coupling is $A_0 = -300$ GeV 
and the common scalar mass $m_0$ is varied in a range between 
50 and 500 GeV. The ratio of the Higgs vacuum expectation 
values is chosen to be $\tan\beta = 4$ and the Higgs mass parameter $\mu$, 
whose absolute value is fixed by electroweak symmetry breaking, is chosen
to be positive. The \rp\ and baryon number violating couplings that involve 
a top (s)quark were taken to be degenerate, 
$ \lp_{312} = \lp_{313} = \lp_{323}  \equiv \lp$.

The signal coming from single stop production in $p\bar{p}$ collisions,
followed by the \rp-conserving decay \st1$\rightarrow b + 
\tilde{\chi}^+_1$, with $\tilde{\chi}^+_1 \rightarrow l + \nu + 
\tilde{\chi}^0_1$, was considered in ref. \cite{berger} together with the 
Standard Model background. The conclusion was that,
for $ 180 < m_{\tilde{t}_1} < 325$ GeV and $\lp > 0.02-0.06$,
it should be possible to discover the top squark at run II of the Tevatron, 
otherwise the limit on the \rp-violating couplings could be lowered to 
$\lp < 0.01-0.03$ at 95\% confidence level. Moreover, existing 
data from run I should allow for a reduction of the limit to $\lp < 0.03-0.2$
for $ 180 < m_{\tilde{t}_1} < 280$ GeV.

We studied \cite{slavich} the bounds on the 
\rp-violating (s)top couplings that can be derived from present experimental
data, and we pointed out that they are in fact more stringent than the 
bound $\lp_{3jk} < 1$ considered in ref. \cite{berger}. 
In particular, we showed that the limits coming from flavor physics put 
severe constraints on the possibility of discovering single top squarks via 
R-parity violation at the Tevatron.

Bounds on the \rp-violating couplings can be derived from \kkb\ mixing 
\cite{masiero,carlson,white}. Flavor-changing neutral currents 
in SUSY models can arise in a ``direct'' way, when the 
flavor violation occurs through flavor violating vertices in the diagrams, 
or in an ``indirect'' way, due to the existence of non diagonal sfermion 
masses in the basis where the fermion masses are diagonal. In minimal 
supergravity scenarios, where the soft mass matrices at the GUT scale are 
flavor diagonal, non diagonal squark masses are generated by flavor violating 
couplings through the Renormalization Group Equations. However, as shown 
in ref. \cite{white}, their contribution to \kkb\ mixing can be neglected.

\begin{figure}
\begin{center}
\epsfig{figure=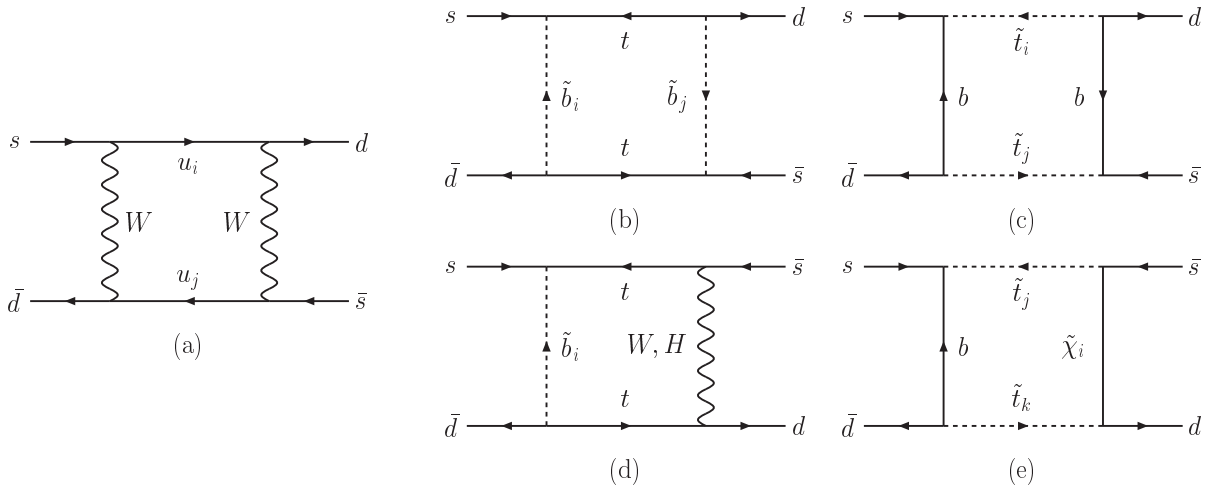,width=16cm}
\end{center}
\caption{Standard Model diagram (a) and diagrams with \rp-violating couplings
(b-e) that give the dominant contributions to \kkb\ mixing. 
The arrows indicate the flow of baryon number.}
\label{fig-box}
\end{figure}

The diagrams that give the dominant contributions to \kkb\ mixing 
in the scenario considered by the authors of ref. \cite{berger} are
shown in fig. \ref{fig-box}.
The most general $\Delta S = 2$ effective Lagrangian can be written as:  
\be
\label{lagr-KK}
{\cal L}_{\rm eff}^{\Delta S = 2} =  \sum_{i=1}^5 C_i Q_i + \sum_{i=1}^3 
\tilde{C}_i\tilde{Q}_i  \; ,
\ee
where the four-fermion operators $Q_i$ and $\tilde{Q}_i$ are defined 
as in ref. \cite{ciuchini}. The operators relevant to this analysis are 
$Q_1 = \bar{d}_L^{\alpha} \gamma_{\mu}s_L^{\alpha}\;
\bar{d}_L^{\beta} \gamma^{\mu}s_L^{\beta}$ ($\alpha,\beta$ are color indices), 
coming from the Standard Model diagram (fig. \ref{fig-box}a),
$\tilde{Q}_1 = \bar{d}_R^{\alpha} \gamma_{\mu}s_R^{\alpha}\;
\bar{d}_R^{\beta} \gamma^{\mu}s_R^{\beta}$, 
coming from the diagrams with four $\lp$ couplings (fig. \ref{fig-box}b-c), 
and $Q_4  =  \bar{d}_R^{\alpha} s_L^{\alpha}\;
\bar{d}_L^{\beta} s_R^{\beta},\; Q_5  =  \bar{d}_R^{\alpha}  s_L^{\beta}\;
\bar{d}_L^{\beta}  s_R^{\alpha}$, coming from the diagrams with two CKM and 
two $\lp$ couplings (fig. \ref{fig-box}d-e).
The corresponding coefficients $C_i$ are evaluated at a common scale $M_S$, 
where the supersymmetric particles are integrated out. We computed
the coefficients that are relevant to the case under consideration: 
their explicit expressions are given in the Appendix.

The contribution of the effective Lagrangian ${\cal L}^{\Delta S = 2}$ 
to the $K_S$--$K_L$ mass difference $\Delta m_K$ is related to the matrix 
element $\langle K^0 | {\cal L}_{\rm eff}^{\Delta S = 2} | \bar{K}^0 \rangle$. 
The coefficients $C_i$ must be evolved from the scale $M_S$, which 
is of order of the masses of the supersymmetric particles,  down to
some hadronic scale $\mu_h$ (e.g. 2 GeV) at which the matrix 
element can be evaluated.
Moreover, the long-distance hadronic processes give  contributions to
the matrix elements  $\langle K^0 | Q_i | \bar{K}^0 \rangle$ that cannot
be evaluated perturbatively, and are parametrized by ``bag factors'' $B_i$
(for the explicit definitions see ref. \cite{ciuchini}). We have calculated 
the contribution of the \rp-violating couplings to $\Delta m_K$, using the 
NLO QCD evolution of the coefficients $C_i$ and the lattice calculations
for the $B_i$ presented in ref. \cite{ciuchini}. 
Due to the large uncertainties that affect the theoretical evaluation 
of $\Delta m_K$ in the Standard Model 
(see e.g. ref. \cite{nierste} and references therein), a 
conservative limit on the \rp-violating couplings can be 
derived by requiring that the contribution to $\Delta m_K$ of the diagrams 
shown in fig. \ref{fig-box}b-e is not larger than the experimental value 
$\Delta m_K^{\rm exp} = (3.489 \pm 0.009) \times 10^{-15}$ GeV. 
The resulting upper bounds on $\lp$, in the minimal 
supergravity scenario considered by the authors of ref. \cite{berger} and 
for the same choice of parameters, are of order $\lp < 0.015-0.020$ for a 
light stop mass ranging between 180 and 325 GeV. 
Thus, the discovery of single stop production via R-parity violation at the 
Tevatron turns out to be unlikely.

There are some other processes that allow to set upper limits on the 
\rp-violating couplings $\lp_{3jk}$. The authors of ref. \cite{carlson} studied
the contribution of the \rp-violating terms to the rare decays of the 
$B^+$ meson, namely $B^+ \rightarrow \bar{K}^0 K^+$ and 
$B^+ \rightarrow \bar{K}^0 \pi^+$. In order to reduce the theoretical 
uncertainties, they considered the ratio of the partial width of the 
rare decay to the partial width of the decay $B^+ \rightarrow K^+ \, J/\psi$, 
which proceeds unsuppressed in the Standard Model. The decay 
$B^+ \rightarrow \bar{K}^0 K^+$ was found to give a more stringent limit
on the \rp-violating couplings. Using updated values for the experimental 
upper bound on the rare decay and for the CKM matrix elements, and taking into 
account the mixing in the stop sector, we obtain the upper limit
$\lp < 0.14-0.23$, depending on the light stop mass. This limit is stringent 
enough to disfavor the possibility that any evidence of single stop 
production is found in the analysis of the run I data, but it is weaker
than the bound derived above from \kkb\ mixing.

In summary, we have improved the existing bounds on the \rp-violating 
couplings $\lp_{3jk}$, showing that they are more stringent than 
those assumed in ref. \cite{berger}. From the study of \kkb\ 
mixing we can set the upper limit $\lp < 0.015-0.020$ for the 
minimal supergravity scenario considered in ref. \cite{berger}. 
As a result, the possibility of detecting single top squark production 
via R-parity violation at the Tevatron turns out to be strongly limited. 

As a final remark, we notice that flavor-changing 
processes allow to set limits only on products of two different
$\lp_{3jk}$ couplings. 
For example, the bound from \kkb\ mixing concerns in general the 
combination $|\lp_{313}\lambda^{\prime\prime\,*}_{323}|^{1/2}$.
Thus, the bounds presented in this work could be evaded 
if only one of the couplings is different from zero, or if there is a
strong hierarchy between the different couplings.
However, such a situation would need to be justified in terms of some flavor 
symmetry to be regarded as natural: this point is currently under 
investigation \cite{slavich}.

\section*{Acknowledgements}
The author thanks the organizers of the XXXV Rencontres de Moriond for 
the pleasant atmosphere in which this work was presented. Many thanks also 
go to Fabio Zwirner for helpful discussions and to Guido Martinelli for useful 
communications.

\section*{Appendix}
We have calculated the contributions to \kkb\ mixing coming from the 
diagrams with \rp-violating (s)top couplings $\lambda^{\prime\prime}_{3jk}$. 
The calculation has been performed in the basis where the quark masses are 
diagonal, and all the flavor changing squark mass insertions have been 
neglected. We have also checked that in the minimal supergravity scenario 
considered in ref. \cite{berger} the contributions coming from MSSM 
diagrams with 
quarks and charged higgs or squarks and charginos are negligible. 
 The coefficients $C_i$ that appear in eq. (\ref{lagr-KK}) are:
\bea
C_1 & = & \!\sum_{i,j=1}^3 \;\frac{g^4}{128 \pi^2}\,
K^*_{i1}\,K_{i2}\,K^*_{j1}\,K_{j2}\, m^2_{u_i}\,m^2_{u_j}\,
\biggr[ I_0 + 2\, I_2/m_W^2 + I_4/4\,m_W^4 \biggr]
(m^2_{u_i},m^2_{u_j},m_W^2,m_W^2) \nn\\ 
\label{coeff1}
&&\\
\tilde{C}_1 & = & \!\sum_{i,j=1}^2 \;\frac{1}{4 \pi^2}\,
|\lambda^{\prime\prime}_{313}\,\lambda^{\prime\prime\,*}_{323}|^2\,
\biggr[
(O^t_{i2} O^t_{j2})^2 \, 
I_4(m_b^2,m_b^2,m^2_{\tilde{t}_i},m^2_{\tilde{t}_j}) 
+ (O^b_{i2} O^b_{j2})^2 \, 
I_4(m^2_{\tilde{b}_i},m^2_{\tilde{b}_j},m^2_{t},m^2_{t})
\biggr]\nn\\
\label{coeff1t}
&&\\
C_5 & = & \sum_{i=1}^2 \;\;\;\frac{g^2}{4 \pi^2}\,
\lambda^{\prime\prime}_{313}\,\lambda^{\prime\prime\,*}_{323}\,
(O^b_{i2})^2 \, K^*_{31}\, K_{32} \, m^2_t \, 
\biggr[ I_2(m^2_{\tilde{b}_i},m_W^2,m^2_t,m^2_t)\, + \nn\\ 
&& \hspace{2.5cm}  \frac{1}{4\, m_W^2}\, 
I_4(m^2_{\tilde{b}_i},m_W^2,m^2_t,m^2_t)\, +\, 
\frac{1}{4\, m_W^2 \tan^2\beta}\,
I_4(m^2_{\tilde{b}_i},m_{H^{+}}^2,m^2_t,m^2_t)\,  \biggr] \nn\\  
& + & \!\!\! \sum_{i,j,k=1}^2 \; \frac{g^2}{8 \pi^2}\,
\lambda^{\prime\prime}_{313}\,\lambda^{\prime\prime\,*}_{323}\,
O^t_{j2}\,O^t_{k2} \, K^*_{31} \, K_{32} \,
\left[V^*_{i1}\, O^t_{j1} - \frac{m_t}{\sqrt{2}\, m_W \sin\beta} \,
V^*_{i2} \, O^t_{j2} \right] \times \nn\\
\label{coeff5}
&& \hspace{2.5cm}
\left[V_{i1}\, O^t_{k1} - \frac{m_t}{\sqrt{2}\, m_W \sin\beta}\, 
V_{i2}\, O^t_{k2} \right] \; 
I_4(m^2_b,m^2_{\tilde{\chi}^{+}_i},m^2_{\tilde{t}_j},m^2_{\tilde{t}_k})
\eea
and $C_4 = - C_5$. In the above equations, $K_{ij}$ are the CKM 
matrix elements, $O^t_{ij}$ and $O^b_{ij}$ are the left-right mixing 
matrices of the stop and sbottom sectors, and $V_{ij}$ is the mixing matrix 
of positive charginos as defined in ref. \cite{haberkane}. 
The masses of the supersymmetric particles and the mixing angles at the
electroweak scale have been calculated with ISAJET \cite{isajet}, and the 
common scale $M_S$ has been chosen as the geometrical 
mean of squark and chargino masses. The functions $I_n$ result from 
integration over the Euclidean momentum $\bar{k}$ of the four particles 
circulating in the loop:
\be
\label{def-I}
I_n(m^2_1,m^2_2,m^2_3,m^2_4) = \int_0^{\infty}
\frac{\bar{k}^n \; d\bar{k}^2}
{(\bar{k}^2+m_1^2)(\bar{k}^2+m_2^2)(\bar{k}^2+m_3^2)(\bar{k}^2+m_4^2)}
\ee

\end{document}